\pdfoutput=1
\documentclass[11pt]{article}
\usepackage{mathfin-paper}

\title{The Fundamental Theorem of Asset Pricing,\\ Formalized in Lean~4}
\author{Raphael Coelho\thanks{Independent researcher. ORCID
  \href{https://orcid.org/0009-0001-6601-1023}{0009-0001-6601-1023}.
  Correspondence: \href{mailto:raphaelrrcoelho@gmail.com}{\texttt{raphaelrrcoelho@gmail.com}}.
  Artifact: \url{https://github.com/formal-applied-math/formal-mathfin}, release \texttt{v1.11.0}.}}
\date{November 2026}

\hypersetup{pdftitle={The Fundamental Theorem of Asset Pricing, Formalized in Lean 4},
  pdfauthor={Raphael Coelho}}

\begin{document}
\maketitle

\begin{abstract}
\noindent
The Fundamental Theorem of Asset Pricing states that a market is free of arbitrage exactly when it
admits an equivalent martingale measure. We formalize it in Lean~4 over Mathlib in three settings: a
finite-state market with one risky asset over a finite horizon (Harrison--Pliska), and one-period markets
with one asset or finitely many assets on an arbitrary probability space. The finite case is the geometry
of a separating hyperplane. In the one-period case the proof builds the equivalent martingale
measure from a minimiser of the potential $\theta\mapsto\E[\log(1+e^{\langle\theta,Y\rangle})]$: absence of
arbitrage makes the potential coercive off the redundant directions, its first-order condition is the
martingale property, and the logistic weight of the minimiser, normalised, is the density; a bounded
tempering removes integrability, so the return need only be measurable. The construction uses no
Hahn--Banach theorem, no $L^0$-closedness argument, no measurable selection and no non-redundancy
hypothesis. One predicate serves every time index, with one forward theorem. The same separation also yields the Artzner--Delbaen--Eber--Heath representation of coherent risk
measures on a finite state space. We add the second theorem on finite spaces (complete if and only if
the martingale measure is unique) and, on any probability space, the reduction of the
Dalang--Morton--Willinger theorem to its conditional one-period form. That conditional step, which
needs a measurable selection, is assumed, not proved, so the multi-period theorem remains open; in continuous time
only the forward direction, for simple strategies, is proved. Every theorem discussed is \texttt{sorry}-free and
uses only Lean's three standard axioms, enforced by the build.
\end{abstract}

\section{Introduction}

A market offers an \emph{arbitrage} if some admissible strategy turns no initial wealth into a payoff
that is non-negative almost surely and strictly positive with positive probability. The Fundamental
Theorem of Asset Pricing identifies the absence of such opportunities with a single analytic object: a
probability measure $Q$, equivalent to the physical measure $P$, under which discounted asset prices
are martingales. No arbitrage holds if and only if such an \emph{equivalent martingale measure}
exists. The equivalence was made precise by Harrison and Kreps~\cite{HarrisonKreps1979} and Harrison
and Pliska~\cite{HarrisonPliska1981}, established for general discrete-time models by Dalang, Morton
and Willinger~\cite{DMW1990}, and carried to continuous time in the no-free-lunch form of Delbaen and
Schachermayer~\cite{DelbaenSchachermayer1994}. It is the result on which arbitrage-free pricing rests:
a price is the $Q$-expectation of a discounted payoff.

The theorem is also a natural test for formal mathematics, because its content is analytic rather than
computational. In the finite case it hides a separating-hyperplane argument; in general it asserts the
existence of a change of measure that is usually obtained non-constructively, by separating a cone in
$L^1$ with a Hahn--Banach functional after proving the cone closed in $L^0$. Fixing exactly which
hypotheses each direction needs, and checking the measure-theoretic details against a kernel, is a task
at which formalization is decisive.

Formalized probability has grown substantially, and with it the measure theory on which pricing
depends. Echenim, Guiol and Peltier verified Cox--Ross--Rubinstein pricing in
Isabelle/HOL~\cite{EchenimGuiolPeltier2020}, Keskin formalized discrete- and continuous-time
martingales~\cite{Keskin2023}, Avigad, H\"olzl and Serafin the central limit
theorem~\cite{AvigadHolzlSerafin2017}, and Degenne and collaborators constructed Brownian motion in
Lean~\cite{DegenneEtAl2025,BrownianMotionRepo}. The nearest precedent for the present paper is
Nagy's Lean~4 derivation from It\^o's lemma to Black--Scholes~\cite{Nagy2026}, which contains the
theorem for a one-period market with one asset and two states. A search in October 2026 of the Lean
ecosystem, the Isabelle Archive of Formal Proofs and the Coq/Rocq ecosystem found no other machine-checked
proof beyond that case: none on an arbitrary probability space, none with several assets on an
arbitrary space, and none over several periods.

\paragraph{What this paper does}
We formalize the theorem in three settings of increasing generality and, in the most general of them,
give the equivalent martingale measure by a formula.
\begin{enumerate}[label=(\arabic*)]
\item \emph{Finite states, finitely many periods} (\cref{sec:finite}). For one risky asset, the backward
  direction is the geometry of a hyperplane separating the attainable gains from the standard simplex
  (\lean{MathFin.ftap_discrete}); in one period the same separation handles finitely many assets.
\item \emph{One period, arbitrary probability space} (\cref{sec:scalar,sec:vector}). For integrable
  returns the proof takes as density of the equivalent martingale measure the normalised logistic weight of
  a minimiser of the softplus potential $\E[\log(1+e^{\langle\theta,Y\rangle})]$; a bounded tempering removes
  integrability. Redundant assets are absorbed rather than excluded, and the return need only be measurable
  (\lean{MathFin.OnePeriodVector.ftap_one_period_vector}). The case of one asset is the instance
  with prices in $\R$ (\lean{MathFin.OnePeriod.ftap_one_period}).
\item \emph{One separation for pricing and risk} (\cref{sec:separation}). The separation lemma of the
  finite case and the one behind the representation of coherent risk measures on a finite state space
  (\lean{MathFin.coherentRisk_isLUB}) are two faces of one finite-dimensional separation, proved side by side
  from shared helper lemmas.
\item \emph{The second theorem} (\cref{sec:second}). On a finite full-support space with full information at the
  horizon, and under no arbitrage, the market is complete if and only if its equivalent martingale
  measure is unique (\lean{MathFin.ftap_two}). One replication argument serves this theorem and its
  Brownian counterpart (\lean{MathFin.measure_eq_of_replicates}).
\item \emph{A precise boundary} (\cref{sec:boundary}). The multi-period theorem on an arbitrary probability space is
  reduced to a conditional one-period statement, which is assumed and not proved, and the continuous-time theorem is
  formalized only in its forward direction; we say exactly what is proved.
\end{enumerate}

This paper is part of a Lean~4~\cite{deMouraUllrich2021} library of formalized mathematical finance
built on Mathlib~\cite{Mathlib2020}; the library is described in a companion
overview~\cite{CoelhoLibrary} and its It\^o calculus in a second companion~\cite{CoelhoIto}. The
no-arbitrage results developed here (\cref{sec:market,sec:finite,sec:scalar,sec:vector,sec:boundary}) live entirely within
Mathlib's measure theory and convex analysis, on its
\lean{MeasureTheory.Measure.withDensity} change of measure and its finite-dimensional separation theorems; they
do not depend on the library's Brownian-motion layer. Only the Brownian instances of the second theorem
(\cref{sec:second}) and the construction of the Black--Scholes measure by Girsanov's theorem use it.

\paragraph{Changes from the first version}
This version corrects three statements of the first. The two-state case was formalized
earlier~\cite{Nagy2026}, so the claim of priority is narrowed. The $d$-asset density is the softplus
analogue of the Esscher measure, not the Esscher measure itself (\cref{sec:why-softplus}). And only one
direction of the correspondence between no arbitrage and coercivity is proved, so the text no longer
calls them equivalent. The version adds the separation shared with coherent risk measures
(\cref{sec:separation}), the finite-state $d$-asset theorem and the forward directions that border the
development (\cref{sec:finite,sec:boundary}). It follows the library from release \texttt{v1.5.0} to
\texttt{v1.11.0}: a single equivalent-martingale-measure predicate for every time index, discrete or
continuous, with prices in an inner-product space and one forward theorem (\cref{sec:market}); the scalar
one-period theorem as the instance of the $d$-asset theorem with prices in $\R$, replacing a separate direct
construction for one asset (\cref{sec:scalar}); the second theorem on finite spaces with its replication argument
and its Brownian and binomial instances (\cref{sec:second}); the reduction of the theorem of Dalang, Morton and Willinger
to its conditional one-period form, with the assumed step stated exactly (\cref{sec:boundary}); and the artifact
at release \texttt{v1.11.0}, whose kernel replay now completes (\cref{sec:artifact}).

\section{The market and the two easy facts}\label{sec:market}

In each setting a position in the traded assets produces a discounted gain, and an arbitrage is a
position whose gain is non-negative almost surely without being almost surely zero. The settings differ
only in what a position is and how its gain is formed.
\begin{itemize}
\item In the \emph{multi-period} model a filtration $\F=(\F_t)_{t\in\N}$ on $(\Omega,P)$ carries an
  adapted discounted price $S_0,S_1,\dots$ valued in a real inner-product space $F$ (one risky asset is
  $F=\R$), and a horizon $T$ is fixed. A strategy $\varphi$ is predictable, with $\varphi_{t+1}$ an
  $F$-valued function measurable for $\F_t$ ($t<T$), and bounded up to $T$. Its discounted gain is the martingale
  transform $(\varphi\cdot S)_T=\sum_{t<T}\langle\varphi_{t+1},S_{t+1}-S_t\rangle$
  (\lean{MathFin.martingaleTransform}).
\item In the \emph{scalar one-period} model a discounted excess return $Y\colon\Omega\to\R$ is held in
  position $\theta\in\R$, with gain $\theta Y$.
\item In the \emph{$d$-asset one-period} model the return $Y\colon\Omega\to F$ takes values in a
  finite-dimensional real inner-product space $F$ (the $d$-asset market is $F=\R^d$), a constant
  position $\theta\in F$ is held, and the gain is $\langle\theta,Y\rangle$. (In Mathlib the $d$-asset
  space is \lean{EuclideanSpace ℝ (Fin d)}, not the product $\R^d$ with its sup norm.)
\end{itemize}
Write $G$ for the discounted gain of a position. \emph{No arbitrage} says that a gain which is
non-negative is null:
\[
  G\ge 0 \quad P\text{-a.s.} \qquad\Longrightarrow\qquad G=0\quad P\text{-a.s.},
\]
which is the textbook ``no $G\ge0$ with $P(G>0)>0$'' in contrapositive form. The condition is stated
under $P$ alone and names no second measure. An \emph{equivalent martingale measure} is a probability
measure $Q$ with $Q\ll P$ and $P\ll Q$ under which discounted prices are fair. In the one-period
settings $Y$ is $Q$-integrable with $\E_Q[Y]=0$ (vector-valued in the $d$-asset case). In the
multi-period setting the predicate \lean{MathFin.IsEMM}\,$\F\,P\,S\,T\,Q$ asks that the price frozen at the horizon,
$t\mapsto S_{\min(t,T)}$, be a $Q$-martingale for $\F$, which for $t<T$ is
$S_t=\E_Q[S_{t+1}\mid\F_t]$. The predicate is stated for any linearly ordered time index, so the same
definition serves the continuous market of \cref{sec:boundary}; a $Q$-martingale on the whole index is an
equivalent martingale measure at every horizon (\lean{MathFin.IsEMM.of_martingale}). Recording equivalence as the
pair of absolute continuities is what lets a $Q$-almost-sure statement transfer back to $P$.
\Cref{lst:defs} shows the $d$-asset one-period definitions; the other settings differ only in the gain. On a
finite space with trivial initial information, the multi-period predicate at horizon $1$ is the
one-period predicate for the return $S_1-S_0$ (\lean{MathFin.isEMM_one_iff_onePeriod},
\lean{MathFin.noArbitrage_one_iff_onePeriod}).

\begin{leancode}[caption={No arbitrage and the equivalent martingale measure, $d$-asset one-period market (\texttt{Foundations/FTAPOnePeriodVector}).},label=lst:defs]
variable {Ω : Type*} {mΩ : MeasurableSpace Ω}
  (P : Measure Ω) [IsProbabilityMeasure P]
  {F : Type*} [NormedAddCommGroup F] [InnerProductSpace ℝ F]
  [FiniteDimensional ℝ F] [MeasurableSpace F] [BorelSpace F] (Y : Ω → F)

def NoArbitrage : Prop :=
  ∀ θ : F, 0 ≤ᵐ[P] (fun ω ↦ inner ℝ θ (Y ω)) →
    (fun ω ↦ inner ℝ θ (Y ω)) =ᵐ[P] 0

structure IsEMM (Q : Measure Ω) : Prop extends IsEquivProbMeasure P Q where
  int  : Integrable Y Q
  fair : ∫ ω, Y ω ∂Q = 0
\end{leancode}
Here \lean{MathFin.IsEquivProbMeasure}\,$P\,Q$ bundles the three facts that $Q$ is a probability measure and that
$Q\ll P$ and $P\ll Q$.

One direction of the theorem is uniform and elementary, and the library proves it once. If an equivalent
martingale measure $Q$ exists, the gains of a predictable strategy bounded up to $T$ form a $Q$-martingale
started at zero (\lean{MathFin.IsEMM.exists_martingale_gains}): each term
$\langle\varphi_{t+1},S_{t+1}-S_t\rangle$ has zero $\F_t$-conditional mean. A terminal gain $G\ge0$ with
$\E_Q[G]=0$ is zero $Q$-almost surely, hence $P$-almost surely by equivalence, so the strategy is not an
arbitrage. This is \lean{MathFin.IsEMM.noArbitrage}, on an arbitrary probability space and for a price valued in
any complete real inner-product space. The other forward directions are instances of it: the
one-period vector theorem (\lean{MathFin.OnePeriodVector.noArbitrage_of_isEMM}, for measurable $Y$) takes the
horizon-$1$ market with trivial $\F_0$, price $S_0=0$, $S_1=Y$ and constant holding $\theta$; the forward direction for the
finite-state market with several assets goes through the one-period vector theorem; and the continuous-time forward theorem of
\cref{sec:boundary} samples the price at the trading dates of a simple strategy. The content of the theorem is the converse,
which the next sections supply in rising generality.

\section{Finite markets: the geometric theorem}\label{sec:finite}

In a finite-state market the equivalent martingale measure is, almost literally, a separating
hyperplane. Fix a finite probability space $(\Omega,P)$ with full support (a measure with $P\{\omega\}>0$
for every $\omega$), a filtration $\F$, an adapted scalar discounted price $S$ and a horizon $T$. The
discounted gains attainable by predictable strategies form a linear subspace of $\R^\Omega$, the span of the
one-period gains $\ind_A(S_{t+1}-S_t)$ with $t<T$ and $A\in\F_t$ (\lean{MathFin.gainsSubspace}); call it the
\emph{gains subspace}. No arbitrage says exactly that this subspace meets the non-negative cone only at the origin,
equivalently that it is disjoint from the standard simplex of probability vectors on $\Omega$.

The simplex is compact and convex and the gains subspace is closed, being finite-dimensional, so the
geometric Hahn--Banach theorem separates them strictly. The library takes this one level higher, in a
form that serves more than pricing. Its root (\cref{lst:cone}) is stated for a closed convex
\emph{cone} $C\subseteq\R^\iota$ disjoint from the simplex, and produces a strictly positive functional
$q$ that is non-positive on $C$. A subspace is a cone closed under negation, so applied to the gains
subspace the one-sided bound becomes two-sided: $q$ annihilates every attainable gain
(\lean{MathFin.exists_pos_dual_of_disjoint_stdSimplex}, derived from the root). Read as a measure, $q$ makes every
martingale transform have zero expectation.

\begin{leancode}[caption={The cone-separation root (\texttt{Foundations/ConvexDuality}).},label=lst:cone]
theorem exists_pos_separating_of_cone_disjoint_simplex
    {ι : Type*} [Fintype ι]
    (C : Set (ι → ℝ)) (hCconv : Convex ℝ C) (hCclosed : IsClosed C)
    (hCcone : ∀ ⦃x⦄, x ∈ C → ∀ ⦃c : ℝ⦄, 0 ≤ c → c • x ∈ C)
    (hdisj : ∀ v : ι → ℝ, (∀ i, 0 ≤ v i) → ∑ i, v i = 1 → v ∉ C) :
    ∃ q : ι → ℝ, (∀ i, 0 < q i) ∧ ∀ v ∈ C, ∑ i, q i * v i ≤ 0
\end{leancode}

The market uses the subspace case; a genuine cone appears in \cref{sec:separation}. The market reaches the theorem in two steps.
Under no arbitrage the gains subspace misses the simplex
(\lean{MathFin.gains_disjoint_stdSimplex}): a non-negative gain is null, and on a full-support space a null
vector has no positive coordinate, while every point of the simplex has one. The strictly positive
annihilator, normalised to a probability vector, is a measure $Q\sim P$ (\lean{MathFin.exists_isEMM_of_pos_dual}); it
annihilates each generator $\ind_A(S_{t+1}-S_t)$, which is the defining property of the conditional expectation, so $S$ is a
$Q$-martingale up to $T$. With the forward direction of \cref{sec:market} this is the biconditional
(\cref{lst:discrete}).

\begin{leancode}[caption={The finite-state theorem (\texttt{Foundations/FTAPDiscrete}). $\Omega$ is \texttt{Finite} and \texttt{Nonempty} with \texttt{MeasurableSingletonClass}, $\mathcal{F}$ a filtration indexed by $\mathbb{N}$, $P$ a measure, $S:\mathbb{N}\to\Omega\to\mathbb{R}$.},label=lst:discrete]
theorem ftap_discrete [Finite Ω] [Nonempty Ω]
    (hS : StronglyAdapted 𝓕 S) (hP : ∀ ω, 0 < P {ω}) :
    NoArbitrage 𝓕 P S T ↔ ∃ Q, IsEMM 𝓕 P S T Q
\end{leancode}

This is the theorem of Harrison and Pliska, the finite case of Dalang, Morton and Willinger. The
statement is slightly stronger than the textbook one: $P$ need only be a measure of full support on the
finite space, not a probability measure. Strategies are bounded up to $T$ in the definition of
\lean{MathFin.NoArbitrage}, which is no restriction on a finite space. The same
kernel gives the one-period theorem for finitely many assets on a finite state space: the gains
subspace is then the span of the assets' excess-return vectors, and the statement is the biconditional
\lean{MathFin.hasEMM_multi_iff_not_hasArbitrage} (\file{Foundations/FTAPMultiState}), between the existence of
a strictly positive probability vector under which every asset's excess return has mean zero and the absence of a
portfolio with non-negative excess return in every state and positive in some state. The multi-period theorem
is stated for one risky asset; with several assets only the generators of the gains subspace would
change, but that version is not in the library. The restriction to finitely many states is what keeps
the separating functional elementary and the measure a genuine probability vector; the next two
sections give it up.

\section{One period, one asset}\label{sec:scalar}

Leaving finitely many states changes the problem. On an arbitrary probability space the attainable gains
need not form a closed set, and the separation argument of \cref{sec:finite} has nothing compact to
grip. For a single scalar return $Y\colon\Omega\to\R$ the theorem is the instance $F=\R$ of the $d$-asset
theorem of \cref{sec:vector}. The library states it for the scalar notions
(\lean{MathFin.OnePeriod.NoArbitrage}, with positions $\theta\in\R$; \lean{MathFin.OnePeriod.IsEMM} is the vector predicate
at $F=\R$) and proves it by transport: on $\R$ the inner product $\langle\theta,y\rangle$ is the product $\theta y$, which
identifies the two no-arbitrage conditions (\lean{MathFin.OnePeriod.noArbitrage_iff_vector}).

\begin{leancode}[caption={The scalar one-period theorem (\texttt{Foundations/FTAPOnePeriod}); $Y:\Omega\to\mathbb{R}$ on an arbitrary probability space.},label=lst:scalar]
theorem ftap_one_period (hY : Measurable Y) :
    NoArbitrage P Y ↔ ∃ Q, IsEMM P Y Q
\end{leancode}

This is the one-asset case of the one-period theorem of Dalang, Morton and Willinger, in the form found
in Föllmer and Schied~\cite[Ch.~1]{FollmerSchied2016}. It carries no proof of its own: both directions are those of
\cref{sec:vector,sec:market}. Earlier versions of the library proved it by a separate construction for one
asset, a two-region density balanced by an explicit choice of weights; the library has since removed that construction in
favour of the instance, and this paper does not describe it. The worked example of \cref{sec:example} shows the
one-asset measure in two states.

\section{One period, many assets: the variational construction}\label{sec:vector}

The $d$-asset one-period theorem is where the construction leaves the textbook proof furthest behind.
The return is now $Y\colon\Omega\to F$ with $F$ a finite-dimensional inner-product space, and a position
is a constant vector $\theta\in F$. The classical route separates the cone of attainable claims from the
non-negative payoffs in $L^1$ by a Hahn--Banach functional, an argument whose infinite-dimensional
duality and cone-closedness are heavy to formalize. We replace it: the equivalent martingale measure is
the minimiser of one smooth convex function on the finite-dimensional space $F$, and every step is a
finite-dimensional or elementary-integral fact. Throughout, the Lean spelling of the gain
$\langle\theta,Y\rangle$ is \lstinline|inner ℝ θ (Y ω)|.

\subsection{The potential}
For $\theta\in F$ set
\[
  f(\theta)=\E\bigl[\log\bigl(1+e^{\langle\theta,Y\rangle}\bigr)\bigr],
\]
the expected \emph{softplus} of the gain. The softplus $u\mapsto\log(1+e^u)$ is smooth and convex, with
derivative the logistic function $\sigma(u)=e^u/(1+e^u)\in(0,1)$, and it satisfies
\[
  u^+\;\le\;\log(1+e^u)\;\le\;\abs{u}+\log 2 .
\]
The upper bound dominates the integrand by $\norm{\theta}\norm{Y}+\log2$, so $f$ is finite and continuous
whenever $Y$ is integrable; the lower bound will give coercivity. Because $\sigma$ is bounded,
differentiation under the integral sign is dominated by $\norm{u}\norm{Y}$, and the derivative in a direction
$u$ is
\[
  \frac{\dd}{\dd t}\Big|_{t=0} f(\theta+tu)=\E\bigl[\sigma(\langle\theta,Y\rangle)\,\langle u,Y\rangle\bigr],
\]
(\lean{MathFin.OnePeriodVector.hasDerivAt_potential_dir}), the directional form of the gradient $\E[\sigma(\langle\theta,Y\rangle)\,Y]$, the $Y$-average weighted by the
logistic of the current gain (\cref{lst:potential}). The potential is also convex, but the proof never
uses that: existence of a minimiser needs only continuity and coercivity, and the first-order condition
holds at any minimiser.

\begin{leancode}[caption={The softplus potential (\texttt{Foundations/FTAPOnePeriodVector}).},label=lst:potential]
noncomputable def softplus (u : ℝ) : ℝ := Real.log (1 + Real.exp u)
noncomputable def logistic (u : ℝ) : ℝ := Real.exp u / (1 + Real.exp u)
noncomputable def potential (θ : F) : ℝ :=
  ∫ ω, softplus (inner ℝ θ (Y ω)) ∂P
\end{leancode}

\subsection{No arbitrage gives coercivity}
No arbitrage is what makes $f$ attain a minimum. Redundant directions are set aside first. The \emph{gains kernel}
\[
  N=\{\theta\in F:\ \langle\theta,Y\rangle=0\ \ P\text{-a.s.}\}
\]
(\lean{MathFin.OnePeriodVector.gainsKernel}) collects the positions with almost surely zero gain. It is a linear subspace,
and $f$ is constant along it, since adding an element of $N$ changes no gain. On the orthogonal complement $N^\perp$ the picture is
rigid. The positive-gain average $g(\theta)=\E[\langle\theta,Y\rangle^+]$ is continuous and positively
homogeneous, and under no arbitrage it is strictly positive on $N^\perp\setminus\{0\}$: if $g(\theta)=0$
then $\langle-\theta,Y\rangle\ge0$ almost surely, so no arbitrage puts $\theta$ in $N$, while
$N\cap N^\perp=\{0\}$. Its minimum $c$ over the unit sphere of $N^\perp$ is therefore positive, and since
the softplus dominates the positive part,
\[
  f(\theta)\;\ge\;g(\theta)\;\ge\;c\,\norm{\theta}\qquad(\theta\in N^\perp).
\]
No arbitrage makes $f$ coercive on $N^\perp$ (\cref{lst:coercive}). The converse holds as well,
though the theorem does not need it and it is not formalized: if $\theta$ is an arbitrage, then at every
point $\eta$ the derivative of $f$ in the direction $-\theta$ is
$-\E[\sigma(\langle\eta,Y\rangle)\langle\theta,Y\rangle]<0$, so $f$ has no minimiser.

\begin{leancode}[caption={Coercivity from no arbitrage (\texttt{Foundations/FTAPOnePeriodVector}).},label=lst:coercive]
lemma exists_pos_lower_bound (hYint : Integrable Y P) (hNA : NoArbitrage P Y)
    (hNbot : (gainsKernel P Y)ᗮ ≠ ⊥) :
    ∃ c > 0, ∀ θ ∈ (gainsKernel P Y)ᗮ, c * ‖θ‖ ≤ potential P Y θ
\end{leancode}

\subsection{The minimiser and its measure}
Coercivity and continuity give a minimiser $\theta_0$ of $f$ on the closed subspace $N^\perp$, by
compactness of a closed ball (\lean{MathFin.OnePeriodVector.exists_global_min_potential}); because
$f$ is flat along $N$ and $F=N\oplus N^\perp$, the same $\theta_0$ minimises $f$ over all of $F$. At a
global minimiser every directional derivative vanishes, so
\[
  \E\bigl[\sigma(\langle\theta_0,Y\rangle)\,Y\bigr]=0
\]
(\cref{lst:foc}). The weight $z=\sigma(\langle\theta_0,Y\rangle)$ lies strictly between $0$ and $1$: it is
bounded, strictly positive, and makes $Y$ fair. The measure $Q$ with $P$-density $z/\E[z]$ is therefore a
probability measure equivalent to $P$, under which $Y$ is integrable with $\E_Q[Y]=0$. If $N^\perp$ is
trivial, every direction is redundant, $Y=0$ almost surely, and $Q=P$ will do. This is the theorem for integrable
returns (\lean{MathFin.OnePeriodVector.exists_isEMM_of_noArbitrage_integrable}).

\begin{leancode}[caption={The first-order condition (\texttt{Foundations/FTAPOnePeriodVector}).},label=lst:foc]
lemma integral_logistic_smul_eq_zero (hY : Measurable Y) (hYint : Integrable Y P)
    {θ₀ : F} (hmin : ∀ θ, potential P Y θ₀ ≤ potential P Y θ) :
    ∫ ω, logistic (inner ℝ θ₀ (Y ω)) • Y ω ∂P = 0
\end{leancode}

\subsection{Removing integrability}
Integrability is bought by \emph{tempering}. The bounded strictly positive weight $w=(1+\norm{Y})^{-1}$,
normalised by $\kappa=\E[w]\in(0,1]$, defines $\tilde P=w/\kappa\cdot P$, equivalent to $P$, under which
$Y$ is integrable because $\norm{Y}w/\kappa\le\kappa^{-1}$. No arbitrage is an almost-sure notion, so it
survives the passage to the equivalent $\tilde P$; the minimisation of the previous subsection takes place under $\tilde P$, and
the measure it produces is equivalent to $\tilde P$, hence to $P$. In general the density of $Q$ with respect to $P$ is therefore proportional to
$\sigma(\langle\theta_0,Y\rangle)\,(1+\norm{Y})^{-1}$, with $\theta_0$ a minimiser of the potential under $\tilde P$.
The backward direction for measurable $Y$ is \lean{MathFin.OnePeriodVector.exists_isEMM_of_noArbitrage}; with the forward direction it is the
biconditional (\cref{lst:vector}).

Two features set this apart from the textbook proof. No non-redundancy hypothesis is needed: redundant
assets live in $N$, the minimisation takes place on $N^\perp$, and the kernel is absorbed rather than
assumed away. And integrability of $Y$ is not assumed: the return need only be measurable.

\begin{leancode}[caption={The $d$-asset one-period theorem (\texttt{Foundations/FTAPOnePeriodVector}); the context is that of \cref{lst:defs}.},label=lst:vector]
theorem ftap_one_period_vector (hY : Measurable Y) :
    NoArbitrage P Y ↔ ∃ Q, IsEMM P Y Q
\end{leancode}

Where the classical proof invokes the dual of $L^\infty$ and the closedness of a cone in $L^0$, this one
invokes coercivity, a minimum over a sphere, and differentiation under an integral, each already in
Mathlib. The reward is an equivalent martingale measure given by a formula in $\theta_0$,
$\sigma(\langle\theta_0,Y\rangle)$ up to normalisation and tempering, rather than produced by a separating
functional. The minimiser itself is obtained by compactness, not computed; ``explicit'' here means a
formula in $\theta_0$, not a closed form. The formula lives in the proof: the theorem asserts existence,
and what the library states about the formula is the first-order condition of \cref{lst:foc}, from which
the martingale property of $Q$ follows by normalisation.

\subsection{Why the softplus}\label{sec:why-softplus}
Variational proofs of the one-period theorem are not new: Rogers~\cite{Rogers1994} obtains the
equivalent martingale measure from a minimisation argument rather than a separation. What matters for a
formalization is the choice of potential, and the softplus is chosen for what the argument asks of it.
For a convex $C^1$ function $\phi$ in place of the softplus, the proof uses four properties:
\begin{enumerate}[label=(\roman*)]
\item $\phi(u)\ge u^+$, which turns no arbitrage into coercivity;
\item $\abs{\phi(u)}\le a\abs{u}+b$, which makes $\E[\phi(\langle\theta,Y\rangle)]$ finite for integrable $Y$;
\item $\phi'$ bounded, which dominates the derivative under the integral by an $L^1$ function;
\item $\phi'>0$, which makes the weight $\phi'(\langle\theta_0,Y\rangle)$ a strictly positive density.
\end{enumerate}
The softplus has all four, with $\phi'=\sigma\in(0,1)$. The exponential $\phi(u)=e^u$ has (i) and (iv)
but neither (ii) nor (iii): its minimiser gives the Esscher density proportional to
$e^{\langle\theta_0,Y\rangle}$, which needs exponential moments of $Y$, and the bounded tempering
$(1+\norm{Y})^{-1}$ supplies only integrability. The logistic density is thus the softplus analogue of
the Esscher measure, not the Esscher measure itself (the first version of this paper conflated the
two). It is bounded, and it exists whenever $Y$ is integrable.

\begin{remark}[interpretation, not formalized]
The convex conjugate of the softplus is $q\log q+(1-q)\log(1-q)$ on $[0,1]$, minus the binary entropy
$H(q)$. Interchanging the minimum over $\theta$ with the supremum over weights, which convex duality
justifies but we do not formalize, identifies $\min_\theta f(\theta)$ with
$\max\{\E[H(z)]:\ 0\le z\le1,\ \E[zY]=0\}$, attained at $z=\sigma(\langle\theta_0,Y\rangle)$. The density
the construction produces is, up to normalisation, the fair weight of maximal expected binary entropy.
\end{remark}

\section{A worked example}\label{sec:example}

A two-state market makes the construction concrete and shows the gains kernel at work. The example is
computed by hand; it illustrates the Lean argument and is not itself a Lean theorem. Let
$\Omega=\{+,-\}$ with $P(+)=P(-)=\tfrac12$, and take two assets whose discounted excess returns are
\[
  Y_1(+)=a,\quad Y_1(-)=-b\quad(a,b>0),\qquad Y_2=c\,Y_1\quad(c\neq0),
\]
so the second asset is redundant. A position $\theta=(\theta_1,\theta_2)$ has gain
$\langle\theta,Y\rangle=(\theta_1+c\,\theta_2)\,Y_1$, which depends on $\theta$ only through the scalar
$s=\theta_1+c\,\theta_2$. The gains kernel is the line $N=\{\theta:\theta_1+c\,\theta_2=0\}$, and the
potential collapses to a function of $s$ alone,
\[
  f(\theta)=\tfrac12\log(1+e^{sa})+\tfrac12\log(1+e^{-sb}).
\]
Its derivative in $s$ is $\tfrac12\bigl(a\,\sigma(sa)-b\,\sigma(-sb)\bigr)$, strictly increasing in $s$,
negative for $s\to-\infty$ and positive for $s\to+\infty$, so it vanishes at a unique $s_0$. The logistic
weight $z=\sigma(s_0Y_1)$, normalised, is the density, and the first-order condition
$a\,\sigma(s_0a)=b\,\sigma(-s_0b)$ is $\E_Q[Y_1]=0$. In two states that pins the measure to
\[
  Q(+)=\frac{b}{a+b},\qquad Q(-)=\frac{a}{a+b},
\]
the classical risk-neutral probability, which is also the explicit weight of the two-state module
(\lean{MathFin.emmWeightUp}, after~\cite{Nagy2026}). The redundant asset adds no constraint: once $Y_1$ is fair,
$Y_2=c\,Y_1$ is fair automatically, as the minimisation over $N^\perp$ predicts. In higher dimension the
first-order condition is no longer a scalar equation, but the structure is the same: the kernel absorbs
the redundant directions, and for integrable returns the logistic of the minimising gain is the density.

\section{One separation for pricing and risk}\label{sec:separation}

The root of \cref{lst:cone} has a sibling. A single point $x_0$ outside a closed convex cone $C$ that
contains the origin is separated from it by a functional $p$ with $\sum_ip_iv_i\le0$ on $C$ and
$\sum_ip_ix_{0,i}>0$ (\lean{MathFin.exists_separating_of_not_mem_cone}). Both are derived in the same file from
Mathlib's geometric Hahn--Banach theorems through two shared lemmas, a basis expansion of a functional
(\lean{MathFin.functional_eq_sum_single}) and the fact that a functional bounded below on a cone is non-negative on it
(\lean{MathFin.functional_nonneg_on_cone}). The first is the pricing reading of separation; the second is the risk reading.

\paragraph{Risk}
A risk measure $\rho$ of losses on a finite state space (a finite measure $P$ on the states) is \emph{coherent} if it is monotone, translation-invariant,
positively homogeneous and subadditive~\cite{ADEH1999} (\lean{MathFin.IsCoherentRiskMeasure}); a larger loss is worse, and
$\rho$ increases with it. Its acceptance set $\{L:\rho(L)\le0\}$ is then a
closed convex cone (\lean{MathFin.IsCoherentRiskMeasure.acceptance_closed}), and a loss $L$ with $\rho(L)>b$ yields a point outside it,
$L-b$. Separating $L-b$ from the acceptance cone and normalising the separating functional
yields a probability vector that prices every acceptable loss non-positively; letting
$b\uparrow\rho(L)$ gives the representation of Artzner, Delbaen, Eber and Heath
(\cref{lst:risk}): $\rho(L)$ is the least upper bound of the expected losses $\E_q[L]$ over the
representing set.

\begin{leancode}[caption={The coherent-risk representation on a finite state space (\texttt{RiskMeasures/AcceptanceSet}).},label=lst:risk]
def representingSet {ι : Type*} [Fintype ι] (ρ : (ι → ℝ) → ℝ) :
    Set (ι → ℝ) :=
  {q | (∀ i, 0 ≤ q i) ∧ (∑ i, q i = 1) ∧
    ∀ Z ∈ {Y : ι → ℝ | ρ Y ≤ 0}, ∑ i, q i * Z i ≤ 0}

theorem coherentRisk_isLUB {ι : Type*} [Fintype ι] [MeasurableSpace ι]
    [MeasurableSingletonClass ι] {P : Measure ι} [IsFiniteMeasure P]
    {ρ : (ι → ℝ) → ℝ} (hρ : IsCoherentRiskMeasure P ρ) (L : ι → ℝ) :
    IsLUB ((fun q ↦ ∑ i, q i * L i) '' representingSet ρ) (ρ L)
\end{leancode}

So the equivalent martingale measure of a finite market and the representing scenarios of a coherent risk
measure are the same kind of object, produced the same way: a strictly positive (respectively
non-negative) normalised functional that separates a cone. In the library this is more than an analogy.
Both theorems import the one module, and the two separations they rest on, cone from simplex and point
from cone, are proved from the same two helper lemmas; they are siblings, not one lemma applied twice. The statement of \lean{MathFin.coherentRisk_isLUB} is also
restated with Mathlib alone, in the convention of positions rather than losses, and checked against the library by an external tool
(\cref{sec:artifact}).

\paragraph{Superhedging}
Superhedging shows where the separation has not yet been carried. Weak duality needs none: in the finite-state one-period
market, every martingale probability vector, equivalent or not, prices a claim below the cost of any portfolio that dominates it
(\lean{MathFin.emm_le_superReplication}), by a direct calculation. Strong duality, that the superhedging price
equals the supremum of martingale prices, is the separation-based converse. It needs the cone of
super-replicable payoffs to be closed, which for a finitely generated cone is classical but is not
available in Mathlib at the library's pin; Mathlib's cone-separation theorem itself is. Strong duality is
not formalized.

\section{Completeness and uniqueness: the second theorem}\label{sec:second}

The first theorem asserts that an equivalent martingale measure exists; the second concerns when it is unique. A multi-period market is
\emph{complete} (\lean{MathFin.IsCompleteMarket}) if every bounded $\F_T$-measurable claim $H$ is replicated
$P$-almost surely, $H=x+(\varphi\cdot S)_T$, by some initial wealth $x$ and some predictable strategy $\varphi$ bounded up to $T$, the
strategy class of no arbitrage.

\paragraph{One replication argument}
The argument that completeness fixes the pricing measure is stated once, with no market in it
(\lean{MathFin.measure_eq_of_replicates}). Let $\mathrm{Gains}$ be a set of real functions on $\Omega$, let $Q_1,Q_2$ be probability
measures dominated by $P$ under which every $G\in\mathrm{Gains}$ is integrable with mean zero, and let $\mathcal{G}$ be a sub-$\sigma$-algebra. If the
indicator of every $\mathcal{G}$-measurable event $A$ equals $x+G$ $P$-almost surely for some $x$ and some $G\in\mathrm{Gains}$, then
$Q_1(A)=Q_2(A)$ for every such $A$: both measures give $A$ the initial wealth $x$. The statement asserts neither the existence of a hedge
nor of a measure; replication and neutrality are hypotheses, discharged in each market. In the finite market the gains are
the terminal gains of bounded predictable strategies, neutral under every equivalent martingale measure
(\lean{MathFin.IsEMM.integrable_and_integral_gains}); in the Brownian market they are the terminal It\^o integrals.

\paragraph{The finite market}
Complete implies unique holds on any probability space for a price valued in an inner-product space:
two equivalent martingale measures of a complete market agree on $\F_T$
(\lean{MathFin.IsEMM.measure_eq_of_isCompleteMarket}). The converse needs a finite space. If the constants together with the
gains subspace were a proper subspace of $\R^\Omega$, a non-zero functional vanishing on it would perturb the strictly positive weights of the
unique equivalent martingale measure into a second one (\lean{MathFin.isCompleteMarket_of_unique}, through
\lean{MathFin.exists_isEMM_of_pos_dual}). Together they give the second theorem (\cref{lst:two}).

\begin{leancode}[caption={The second theorem on a finite space (\texttt{Foundations/FTAPComplete}); the context is that of \cref{lst:discrete}.},label=lst:two]
theorem ftap_two (hS : StronglyAdapted 𝓕 S) (hP : ∀ ω, 0 < P {ω})
    (hT : 𝓕 T = mΩ) (hNA : NoArbitrage 𝓕 P S T) :
    IsCompleteMarket 𝓕 P S T ↔ ∃! Q, IsEMM 𝓕 P S T Q
\end{leancode}

The hypotheses are those of \cref{lst:discrete} together with full information at the horizon, $\F_T=\mathcal{P}(\Omega)$
(for a finite $\Omega$ with measurable singletons, the whole power set), and no arbitrage, which supplies the existence of an
equivalent martingale measure. The statement is for a finite full-support space and one real discounted price; a $d$-asset version is not claimed.

\paragraph{The binomial tree}
In the $T$-period Cox--Ross--Rubinstein tree on the flips $\Omega=\{\mathrm{false},\mathrm{true}\}^T$ with $S_0>0$, $0<d<u$, $T\ge1$ and a real-world law of
independent Bernoulli$(p)$ flips, $0<p<1$, no arbitrage holds exactly when $d<e^r<u$
(\lean{MathFin.noArbitrage_horizonSpot_iff}); the direction from the inequality is the equivalent martingale measure
together with the forward theorem, and the converse is an explicit first-period strategy. Under $d<e^r<u$ the equivalent martingale measure exists
and is unique (\lean{MathFin.existsUnique_isEMM_horizonSpot}), and the tree is complete if and only if it is unique
(\lean{MathFin.isCompleteMarket_horizonSpot_iff}), the instance of \cref{lst:two} for the tree. The multi-period no-arbitrage statement for the tree
thus goes through the discrete FTAP of \cref{sec:finite}.

\paragraph{The Brownian market}
In the market whose price is driven by a Brownian motion $B$ on $(\Omega,\mu)$ with continuous paths, a probability measure $Q\ll\mu$ under which
every terminal It\^o integral $\int_0^T\varphi\,\dd B$ is integrable with mean zero agrees with $\mu$ on $\F^B_T$
(\lean{MathFin.measure_eq_of_pricesGainsAtZero}). This is the replication argument for the class of terminal It\^o integrals, applied to $Q$ and to
$\mu$ itself, completeness being the representation of the indicator of every $\F^B_T$-event as its mean plus an It\^o
integral~\cite{CoelhoIto}. For a driftless price $S=S_0+\int\sigma\,\dd B$ with $\sigma\neq0$ almost everywhere, the neutrality of the
gains is itself derived: a probability measure $Q=D\cdot\mu$ with square-integrable density $D\ge0$ under which $S$ is a martingale agrees with $\mu$ on
$\F^B_T$ (\lean{MathFin.PricingMeasureL2Density.measure_eq_of_density}, \file{Foundations/PricingMeasureL2Density}).
The square-integrability of the density is used and is not removed by the argument. These are the direction from completeness to uniqueness; the
converse, which needs the extreme-point characterisation of Jacod and Yor, is not formalized.

\section{What is proved, and what is not}\label{sec:boundary}

The theorems above are complete and machine-checked: the finite-state one-period theorem with several
assets, the finite-state multi-period theorem with one asset, and the one-period theorem on an arbitrary
probability space with one asset and with several. Each is a biconditional between no arbitrage and the
existence of an equivalent martingale measure (\cref{fig:ladder}).

\begin{figure}[tbp]
\centering
\begin{tikzpicture}[
  node distance=7mm and 6mm,
  box/.style={draw=mfaccent!70, fill=mfbg, rounded corners=2pt, align=center,
    text width=64mm, inner sep=4pt, font=\small},
  open/.style={box, dashed, fill=white},
  arr/.style={-{Stealth[length=2.2mm]}, draw=mfaccent},
  lab/.style={font=\footnotesize\itshape, text=mfmuted}]
\node[box] (a) {Finite states, one period, $d$ assets\\[1pt]
  \texttt{\footnotesize hasEMM\_multi\_iff\_not\_hasArbitrage}};
\node[box, below=of a] (b) {Finite states, $T$ periods, one asset\\[1pt]
  Harrison--Pliska $\cdot$ \texttt{\footnotesize ftap\_discrete}};
\node[box, below=of b] (c) {Arbitrary $\Omega$, one period, $d$ assets\\[1pt]
  variational EMM $\cdot$ \texttt{\footnotesize ftap\_one\_period\_vector}};
\node[open, below=of c] (d) {Arbitrary $\Omega$, $T$ periods\\[1pt]
  Dalang--Morton--Willinger $\cdot$ reduction proved,\\ conditional one-period step assumed};
\node[open, below=of d] (e) {Continuous time\\[1pt]
  Delbaen--Schachermayer $\cdot$ forward direction only};
\draw[arr] (a) -- node[right, lab] {add periods} (b);
\draw[arr] (b) -- node[right, lab] {drop finiteness} (c);
\draw[arr, dashed] (c) -- node[right, lab] {add dynamics} (d);
\draw[arr, dashed] (d) -- node[right, lab] {continuous trading} (e);
\end{tikzpicture}
\caption{The settings formalized here (solid) and those that remain open or partial (dashed). Each solid
box is a machine-checked biconditional between no arbitrage and an equivalent martingale measure; the scalar
one-period theorem is the instance of the third box with prices in $\R$.}
\label{fig:ladder}
\end{figure}

\paragraph{The multi-period theorem on an arbitrary space}
The forward direction holds on an arbitrary probability space over several periods
(\lean{MathFin.IsEMM.noArbitrage}, \cref{sec:market}). The theorem of Dalang, Morton and Willinger places a
multi-period market on an arbitrary probability space: trading is dynamic, strategies are predictable
processes, and the attainable gains live in $L^0$ rather than in a finite-dimensional space. Its classical
proofs assemble the one-period conditional measures into a single density across time, and need a measurable selection, or the
closedness of the cone of attainable claims in $L^0$, to produce each conditional measure~\cite{DMW1990,KabanovStricker2001}.
The library proves everything around that step, on any probability space with a price valued in an inner-product space, and assumes the step itself.
\begin{itemize}
\item \emph{No arbitrage is local.} For a price adapted up to $T$, the multi-period market is arbitrage free if and only if,
  for every $t<T$, the one-period market conditional on $\F_t$ is: every $\F_t$-measurable position $\theta$, bounded or not,
  whose one-period gain $\langle\theta,S_{t+1}-S_t\rangle$ is almost surely non-negative has an almost surely zero gain
  (\lean{MathFin.noArbitrage_iff_noArbitrageLocal}). The direction towards the one-period markets truncates $\theta$ to the bounded
  position $\theta\ind_{\{\norm{\theta}\le n\}}$; the converse is an induction on the horizon.
\item \emph{One-period densities paste.} For an adapted integrable price valued in a complete inner-product space, suppose that each period $t<T$ carries a bounded, strictly
  positive, $\F_{t+1}$-measurable weight $Z_t$ with $P[Z_t\mid\F_t]=1$ and $P[Z_tS_{t+1}\mid\F_t]=S_t$, a one-period
  equivalent martingale measure conditional on $\F_t$. Then $P$ with density $\prod_{t<T}Z_t$ is an equivalent martingale measure for the
  multi-period market (\lean{MathFin.LocalEMMDensities.isEMM_pastedMeasure}); the proof is the tower property, peeling one factor at a time.
\item \emph{The assumed step.} \lean{MathFin.ConditionalOnePeriodFTAP}$\,\F\,P\,S\,T$ is the hypothesis that conditional one-period no arbitrage
  at each period $t<T$ yields such a weight. Under it, for an integrable price valued in a complete inner-product space, no arbitrage gives
  an equivalent martingale measure
  (\lean{MathFin.exists_isEMM_of_noArbitrage_of_conditionalFTAP}), and with the forward direction no arbitrage is equivalent to its
  existence (\lean{MathFin.noArbitrage_iff_exists_isEMM}). The corpus entry for this composition is a \status{reduced\_core} entry: it is not the
  theorem of Dalang, Morton and Willinger, whose substance is the hypothesis.
\end{itemize}
The hypothesis is the conditional counterpart of the one-period theorem of \cref{sec:vector}. That theorem, with trivial $\F_t$, produces a fair
measure for one period without conditioning, but the library does not derive the hypothesis from it: conditioning on $\F_t$ requires choosing
the minimiser $\theta_0$ of the softplus potential measurably in the state, and this is where a measurable selection re-enters; it is not attempted
here. The hypothesis is the one-period theorem of the literature only for a finite-dimensional price space; for an infinite-dimensional space the
statement under it may be vacuous. So the multi-period theorem of Dalang, Morton and Willinger is open in the library apart from this reduction. So is
the finite-state multi-period theorem with several assets. At the level of contracts the hypothesis yields a consequence, again a reduced core: for
an adapted integrable price with $S_0$ constant, no arbitrage and the hypothesis give one equivalent martingale measure under which, for every
strike and maturity, the forward is worth $S_0-K$ and call minus put equals $S_0-K$
(\lean{MathFin.Contracts.DiscreteMarket.exists_isEMM_parity}, valuation by \lean{MathFin.Contracts.Contract.value}~\cite{CoelhoLibrary}).

\paragraph{Continuous time}
In continuous time the library proves the forward direction for simple strategies: if $Q\sim P$ and the
discounted price, valued in a finite-dimensional inner-product space, is a $Q$-martingale up to $T$, then no strategy with finitely many trading dates and bounded
predictable holdings is an arbitrage (\lean{MathFin.ContinuousMarket.isEMM_noArbitrageSimple}, \file{Foundations/ContinuousMarket}). The proof has no argument of
its own: a simple strategy is a discrete one for the price sampled at its trading dates, and the discrete forward theorem applies.
For the Black--Scholes model the equivalent martingale measure is not assumed but constructed, by a
Girsanov change of measure: under the physical measure, with market price of risk $\theta=(\mu-r)/\sigma$, the measure with density $\exp(-\theta B_T-\tfrac12\theta^2T)$ is an
equivalent martingale measure up to $T$ for the discounted price (\lean{MathFin.bs_discounted_isEMM}~\cite{CoelhoLibrary,CoelhoIto}).
The converse, that no free lunch with
vanishing risk yields an equivalent ($\sigma$-)martingale measure~\cite{DelbaenSchachermayer1994}, is not
formalized. Of the second theorem only the direction from completeness to uniqueness is formalized in the Brownian market (\cref{sec:second}).

\section{The artifact}\label{sec:artifact}

The development is part of a Lean~4 library that builds against pinned versions of its dependencies:
the Lean toolchain \texttt{v4.35.0}, Mathlib at revision \texttt{e91b128}, and the
\texttt{BrownianMotion} package~\cite{BrownianMotionRepo} at \texttt{497e5e2}, which the no-arbitrage results of this
paper do not use. The state described here is release \texttt{v1.11.0} of November 2026,
archived on Zenodo~\cite{Zenodo}. A plain \texttt{lake build} from a clean checkout is the canonical
check.

Honesty about what the kernel accepts is enforced rather than asserted. Two files run
\texttt{\#print axioms} under \texttt{\#guard\_msgs}: a curated audit of headline results, and a generated one
that pins every library constant a benchmark proof cites, resolved by declaration rather than by name.
Each theorem of \cref{tab:results} depends only on \texttt{propext}, \texttt{Classical.choice} and
\texttt{Quot.sound}; a \texttt{sorry} or an additional axiom anywhere beneath one of them fails the build. The
reduced-core entry of the table carries its assumed hypothesis in the statement, so that the axiom audit does not hide it.
A verification ledger records, for every benchmark entry, a hash of the exact sources and toolchain pins it
was last checked against, so that an entry whose inputs change cannot keep a stale verdict. A kernel replay
(\texttt{leanchecker}) of every content module, below the elaborator, runs monthly in six parallel shards and completes.
For the coherent-risk theorem of \cref{sec:separation} the repository also carries a Mathlib-only restatement,
paired with a witness that its hypothesis is satisfiable, and a proof of both from the library. Lean's comparator,
\texttt{lake comparator}~\cite{Comparator}, run with \texttt{--paranoid}, checks that the proved statements are exactly the restated ones, that only the
three standard axioms are used, and that every external kernel bundled with Lean accepts the proof. The results discussed are collected in
\cref{tab:results}.

\begin{table}[tbp]
\centering\footnotesize
\begin{tabularx}{\linewidth}{@{}>{\raggedright\arraybackslash}p{0.43\linewidth}>{\raggedright\arraybackslash}X>{\raggedright\arraybackslash}p{0.27\linewidth}@{}}
\toprule
Lean declaration & Module & Benchmark entry \\
\midrule
\lean{MathFin.ftap_discrete} & \file{Foundations/FTAPDiscrete} & \bench{mf-ftap-discrete-complete} \\
\lean{MathFin.hasEMM_multi_iff_not_hasArbitrage} & \file{Foundations/FTAPMultiState} & \bench{mf-ftap-single-period-complete} \\
\lean{MathFin.OnePeriod.ftap_one_period} & \file{Foundations/FTAPOnePeriod} & \bench{mf-ftap-one-period-general} \\
\lean{MathFin.OnePeriodVector.ftap_one_period_vector} & \file{Foundations/FTAPOnePeriodVector} & \bench{mf-ftap-one-period-vector} \\
\lean{MathFin.exists_pos_separating_of_cone_disjoint_simplex} & \file{Foundations/ConvexDuality} & \bench{mf-convex-duality-root} \\
\lean{MathFin.coherentRisk_isLUB} & \file{RiskMeasures/AcceptanceSet} & \bench{mf-coherent-risk-representation} \\
\lean{MathFin.emm_le_superReplication} & \file{Foundations/SuperhedgingDuality} & \bench{mf-superhedging-emm-bound} \\
\lean{MathFin.IsEMM.noArbitrage} & \file{Foundations/FTAP} & \bench{mart-thm-2.6.7} \\
\lean{MathFin.ftap_two} & \file{Foundations/FTAPComplete} & \bench{mf-ftap-two-discrete} \\
\lean{MathFin.measure_eq_of_replicates} & \file{Foundations/ReplicationPricing} & \bench{mf-replication-determines-pricing-measure} \\
\lean{MathFin.noArbitrage_horizonSpot_iff} & \file{Binomial/MultiPeriodFTAP} & \bench{mf-binomial-multi-period-no-arbitrage-iff} \\
\lean{MathFin.noArbitrage_iff_noArbitrageLocal} & \file{Foundations/NoArbitrageLocal} & \bench{mf-no-arbitrage-local} \\
\lean{MathFin.LocalEMMDensities.isEMM_pastedMeasure} & \file{Foundations/DensityPasting} & \bench{mf-density-pasting-emm} \\
\lean{MathFin.noArbitrage_iff_exists_isEMM} (assumes \lean{MathFin.ConditionalOnePeriodFTAP}) & \file{Foundations/DalangMortonWillinger} & \bench{mf-dmw-conditional} (\status{reduced\_core}) \\
\lean{MathFin.ContinuousMarket.isEMM_noArbitrageSimple} & \file{Foundations/ContinuousMarket} & \bench{gir-continuous-emm-forward} \\
\lean{MathFin.PricingMeasureL2Density.measure_eq_of_density} & \file{Foundations/PricingMeasureL2Density} & \bench{gir-pricing-measure-density} \\
\bottomrule
\end{tabularx}
\caption{The results discussed in this paper, their modules under \texttt{MathFin/}, and their benchmark
entries; every entry is \status{full} but the one marked. Every declaration is pinned by the axiom audits to \texttt{propext}, \texttt{Classical.choice} and
\texttt{Quot.sound}.}
\label{tab:results}
\end{table}

Within the wider library these sit among 804 catalogued results: 766 complete formalizations, 18 thin
wrappers over Mathlib or \texttt{BrownianMotion} theorems, and 20 reduced cores that prove less than the
result they are named after, with the 784 complete-or-wrapper entries counting as delivered. The catalogue
is published as a dataset in which every entry carries its statement, its status and a description of
what its formalization leaves out~\cite{HFDataset}. The library is released under the Apache~2.0
licence.

\paragraph{Provenance}
The Lean development was written by the author with an AI coding assistant (Claude Code), as recorded in
the repository's provenance file; every proof is checked by the Lean kernel, and the gates above apply
to it unchanged. This manuscript was revised with the same assistant, and its statements were checked
against the Lean sources of release \texttt{v1.11.0}.


\end{document}